\documentclass[twocolumn,showpacs,preprintnumbers,amsmath,amssymb]{revtex4}

\usepackage{graphicx}
\usepackage{dcolumn}
\usepackage{bm}
\usepackage{verbatim}
\usepackage{subfigure}
\usepackage{rotating}
\usepackage{afterpage}
\usepackage[
unicode=true,%
colorlinks=true,%
linkcolor=blue,%
urlcolor=blue]{hyperref}

\newcommand{\df}{\Delta \phi}

\newcommand{\na}{\langle n^A \rangle}
\newcommand{\nb}{\langle n^B \rangle}
\newcommand{\ja}{j^A}
\newcommand{\jb}{j^B}
\newcommand{\ba}{b^A}
\newcommand{\bb}{b^B}
\newcommand{\same}{\frac{d\langle n_{same}^{AB} \rangle}{d\Delta\phi}}
\newcommand{\mix}{\frac{d\langle n_{mix}^{AB} \rangle}{d\Delta\phi}}
\newcommand{\jet}{\frac{d\langle n_{jet}^{AB} \rangle}{d\Delta\phi}}
\newcommand{\bg}{\frac{d\langle n_{bg}^{AB} \rangle}{d\Delta\phi}}

\begin{document}
\preprint{APS/123-QED}

\title{Extraction of Correlated Jet Pair Signals in Relativistic Heavy Ion Collisions}
\author{Anne Sickles}
\email{anne@bnl.gov}
\affiliation{
Department of Physics
Brookhaven National Laboratory
Upton, NY 11973
}

\author{Michael P. McCumber}
\affiliation{
Department of Physics and Astronomy
SUNY Stony Brook
Stony Brook, NY 11794
}

\author{Andrew Adare}
\affiliation{
Department of Physics
University of Colorado
Boulder, CO 80309
}

\date{\today}

\begin{abstract}
  Multi-particle correlation techniques are frequently used to study
  jet shapes and yields in hadronic and nuclear collisions. To date, a  
  standard assumption applied in such analyses is that the observed
  correlations arise from either jets and associated hard scattering
  phenomena, or from a background component due to 
  combinatorial pairs connected only through whole even correlations. Within this assumption of two
  essentially independent sources, a fundamental problem centers
  around determining the relative contributions of each component. We
  discuss the methods commonly used to establish the background yield 
  in jet correlation analyses, with a full explanation of the absolute
  background normalization technique which establishes the background
  yield without assumptions about the shape of jet correlations.  This
  is especially important in relativistic heavy ion collisions where
  the jet shapes are significantly distorted from the well separated
  back-to-back di-jets observed in proton-proton collisions.
\end{abstract}

\maketitle

\section{Introduction} \label{sec:intro} 

One of the main goals of high energy nuclear physics experiments 
is to produce and study the properties of hot and dense nuclear matter
in ultra-relativistic heavy ion collisions. Probing this matter with
strongly interacting partons produced in hard scattering processes in
the initial stages of the collision has been a useful tool for
deducing the properties of the created matter~\cite{brahmswp, phoboswp, starwp,ppg048}.
In high energy physics, hard processes are often studied using an
algorithm to identify and reconstruct jets composed of partonic
fragments to infer properties of the parent parton. In heavy ion
collisions, such measurements are extremely difficult due to the large
number of total particles compared to the multiplicity within a
jet. Instead, one-, two- and three-particle observables have been used
to study hard scattering phenomena. The study of particle yields and
their correlations, when compared to expectations from $p$+$p$
collisions, provides quantitative information about the modification
of jet production in the presence of hot nuclear matter.

Two particle correlations have been a particularly important tool for
describing modification of jets in the nuclear environment. Dramatic
modification has been observed in jet yields and shapes in central
heavy ion collisions compared to the smaller baseline colliding
systems. In $p$+$p$ collisions, two particle azimuthal correlations
have a small angle near-side peak from particles which arise from the
same jet and an away-side peak at $\Delta\phi=\pi$ from particles
which arise from opposing sides of a pair of back-to-back di-jets. In
central heavy ion collisions at $\sqrt{s_{NN}}=$ 20-200 GeV the
picture is qualitatively different. The away-side peak is shifted
aside to $\Delta\phi\approx2$~rad and is much broader than in
$p$+$p$ collisions~\cite{ppg067} while the near-side peak also
contains an elongated structure in the longitudinal
direction~\cite{starlowpt,Alver:2009id}. However, quantification of
these results requires removal of the combinatorial background of
particle pairs which are only correlated through whole-event
properties. This combinatorial background is large and must be removed
in order to extract correlations arising from jet processes. The
background is largest in central events and at lower $p_{T}$, where it
can be as much as two orders of magnitude larger than the jet
signal.

As the statistical precision of the correlations data from relativistic 
heavy ion collisions increases, better understanding of the accuracy and
limitations of the experimental methods used to measure two particle
correlations is needed. To this end we discuss the merits and
drawbacks associated with methods commonly used to extract the jet
signal in multi-particle angular correlations and specifically discuss
a method to determine the combinatorial background that requires no
assumptions about the shape of the jet correlations.

\section{Two Particle Angular Correlations}
The angular correlation technique has been used extensively to deduce
jet properties in hadronic and nuclear collisions, and is described in
detail in several places \cite{starlowpt,ppg032,ppg083}. A brief
description of the method is provided here. For simplicity, we will
focus the following discussion on two-particle azimuthal correlations
projected onto the transverse plane.

The angular correlation technique statistically examines the
relationship between particles classified as triggers (denoted as type
$A$) and partners (denoted as type $B$). When studying jets, trigger
particles are typically selected to have larger $p_{T}$ values than
partners and both categories usually have $p_{T}>$ 1 GeV/$c$, though this
is necessary.

Due to the back-to-back production of partons by a hard scattering
process, the distribution of relative azimuthal angles $\df = \phi^{A}
- \phi^{B}$ is expected to peak at $\df=0$ and $\pi$. However, the
measured shape of this distribution may be significantly distorted when
measured in a real detector due to non-uniform angular coverage and
dead or inefficient areas. The shape of the detector acceptance in
$\df$ can be determined by pairing triggers and partners from
different events through a process of event mixing. A mixed pair 
contains correlations due to the detector acceptance but the physical
correlations are eliminated. The ratio of the same event pair
distribution to that of mixed pairs allows pair acceptance effects to
cancel, leaving a distribution reflecting only physical
correlations. This ratio is conventionally defined as the correlation 
function, $C(\df)$:  
\begin{eqnarray} \label{eq:defC}
  C\left(\Delta\phi\right) 
  =  \frac{\displaystyle \same }{\displaystyle \mix }
  \frac{\displaystyle \int{ \mix d\Delta\phi} }
  {\displaystyle \int{ \same d\Delta\phi} } 
\end{eqnarray}
where $n^{AB}$ is the number of pairs per event for either same or
mixed pairs
and $\langle~\rangle$ indicates averaging over many events within some
centrality selection. 
\footnote{
  Though other definitions exist, this choice permits the interchange of
  corrected and measured distributions. 
}.

\begin{figure}[tb]
\centering
\includegraphics[width=0.75\linewidth]{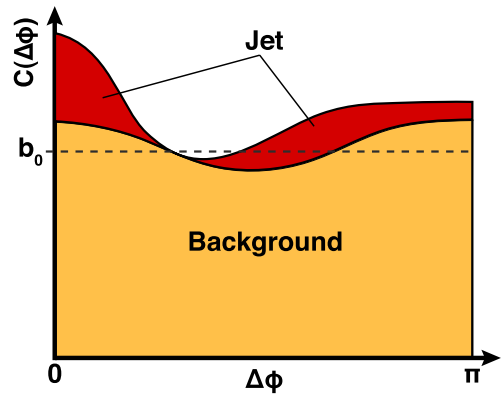}
\caption{(Color online) An illustration of the role of $b_0$ in the two source model separation of jet
  and background pairs.}
\label{b0_cartoon}
\end{figure}

Under the assumption that observed correlations arise from two
independent and separable sources, the $d\langle n^{AB}_{same} \rangle/d\df$ distribution
consists of $AB$ pairs that are correlated with each other by only whole event
correlations (which we call ``background''
pairs), and those where the particles carry additional spatial correlation.
The source of these additional correlations is generally thought of as
an association to a particular hard scattering, thus these pairs are
labelled ``jet'' pairs
\footnote{ Other sources of correlations such as HBT and resonance
  decays are small at the momenta used to study jets. }.   
Single particles are assumed to be either from jets or some non-jet
source.  
Background pairs contain pairs where 1) one particle is from a jet and
the other is not, 2) pairs where both particles are not from jets, and 
3) pairs where both particles are from jets and not associated with the same
hard scattering.
\begin{eqnarray} \label{eq:decomp}
   \same = \jet + \bg
\end{eqnarray}
The {\it two source model} can be written as: 
\begin{eqnarray} \label{eq:2src}
  C\left(\df\right) = J\left(\df\right) +b_0 \left(1+2\langle v_2^Av_2^B\rangle \cos
  \left(  2\df \right) \right).
\end{eqnarray} 
The quadrupole anisotropy coefficients $v_2^{\{A, B\}}$ are taken as inputs from
independent measurements of type $A$ and $B$ particles. Higher order
terms of the anisotropy expansion are smaller and often neglected. In Eq~\ref{eq:2src}, the
approximation is made that the background pair anisotropy is equivalent to 
the product of the single-particle anisotropy coefficients:
\begin{eqnarray}
  \label{eq:v2fac}
  \langle v_2^A v_2^B \rangle = \langle v_2^A \rangle \langle v_2^B
  \rangle. 
\end{eqnarray}
Within the
assumption that the azimuthal modulation of the background is independent of the
jet signal, the fundamental problem of decomposing the correlation
function into jet and background components amounts to the
determination of the background level, $b_0$, as shown in
Fig~\ref{b0_cartoon}.
By equating the final background terms in Eq~\ref{eq:decomp}
and \ref{eq:2src} and integrating over $\df$, using the definition of
$C(\df)$ in Eq~\ref{eq:defC}, the background level can be expressed
in terms of per-event pair multiplicities as:
\begin{eqnarray}\label{eq:b0def}
  b_0 &=& \frac{\langle n^{AB}_{bg} \rangle}{\langle n^{AB}_{same} \rangle}.
\end{eqnarray}
In practice $b_0$ and thereby $\langle n^{AB}_{bg} \rangle$ has been calculated using
two approaches: the first involves scaling $b_0$ so that some portion
of the angular distribution is zero, while the second uses a
calculation to obtain the quantities in Eq~\ref{eq:b0def}. We follow
the conventional nomenclature and refer to the former class of methods
as the Zero Yield At Minimum (ZYAM)~\cite{sbzyam} approach, and the
latter as the Absolute Background Subtraction
(ABS)~\cite{ppg033,ppg067,ppg083} technique. These methods are
discussed in detail in the following sections. 

A quantity that is frequently of interest is the per-trigger jet pair
yield, which describes the conditional jet pair multiplicity as a
function of relative azimuthal angle. The term ``conditional'' refers
to the coincidence of a trigger-partner pair within some angular
region divided by the production rate of triggers in the same centrality
category. It can be shown that the per-trigger yield of total pairs
(i.e. including jets and background) is related to the correlation
function in a simple way:
\begin{eqnarray}\label{eq:sumrule}
  \frac{1}{\na} \same = \frac{\langle n^{AB}_{same} \rangle}{\na} \frac{C(\df)}{\int
    C(\df^{\prime})d\df^{\prime}}
\end{eqnarray}
where $\na$ is the mean number of triggers per event. Since $J(\df)$
is the fraction of the correlation function from jets, the per-trigger
yield of the jet pairs can be calculated as in:
\begin{eqnarray}
\frac{1}{\na} \jet = \frac{\langle n^{AB}_{same} \rangle}{\na} \frac{J(\df)}{\int
  C(\df^{\prime})d\df^{\prime}}.
\end{eqnarray}

\section{The ABS method}
The absolute background normalization method is based on the assumption that
the background is combinatorial in nature and that hard scattering 
results in large correlations between the production rates of jet
particles. The combinatorial background carries only event-wise
correlations. The background pair production rate is given by the
product of the single particle production rates: $\langle n^{AB}_{bg}
\rangle = \na \nb$. Thus the true combinatorial background level in an
ideal case is simply: 
\begin{eqnarray}
b^{ideal}_{0} = \frac{\na \nb}{\langle n^{AB}_{same}\rangle}
\label{eq:b0_abs}
\end{eqnarray}
The values of $\na$ and $\nb$ are measurable and, in the absence of
other correlations, an accurate knowledge of these quantities is
sufficient to determine the background level. However, $n^A$ and
$n^B$ are both dependent on the event centrality, and this
dependence gives rise to a multiplicity correlation. More central
events typically have both larger $n^A$ and $n^B$. Because of the
correlation between $n^A$ and $n^B$ when events are grouped into
centrality bins, the number of measured background pairs is larger than
that expected from Eq~\ref{eq:b0_abs}, $\langle n^{AB}_{bg} \rangle >
\na\nb$. An
additional correction is needed to account for this effect when
calculations are made for data selections that span a finite
centrality range. 

\subsection{Centrality Bias Correction}

\begin{figure}[tb]
\centering
\includegraphics[width=0.75\linewidth]{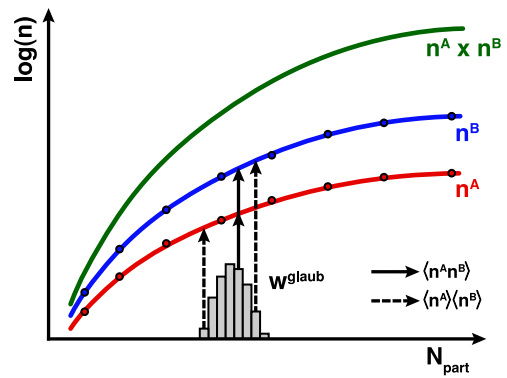}
\caption{(Color online) Schematic of the basic features of the $\xi$ calculation.}
\label{xicorr}
\end{figure}

We define a scale factor correction, $\xi$, for the production rate
product that accounts for the covariance effects arising from the
centrality bias; $\xi$ is defined as: 
\begin{eqnarray}
  \xi \equiv \frac{\langle n^A n^B \rangle}
        {\na \nb }
\end{eqnarray}
The diagram shown in Fig~\ref{xicorr} depicts
the basic features of the procedure to calculate the centrality correction.
The single-particle production rates, $n^A$ and $n^B$, are a function of some
global property of the collision related to particle production, such
as the number of nucleons participating in the collisions ($N_{part}$)
or the number of binary nucleon-nucleon collisions ($N_{coll}$). The
variations of $\na$ and $\nb$ are measured over specific intervals in
these parameters, which are specified by the width of the centrality
bins used for event mixing. Typically centrality ranges are
approximately 5\%. The values of $N_{part}$ or $N_{coll}$ are based on the results
of a Glauber Monte Carlo simulation~\cite{glauber_review}. For the purpose of
narration, we discuss the calculation using $N_{part}$, though
$N_{coll}$ is of equivalent utility, and calculations using this
parametrization are also made in parallel. In practice, the variation in $\xi$
from using the two different parametrizations provides a gauge
of the systematic uncertainty inherent in the method; the two
parameterizations bracket the expected scalings of hard and soft
production that may both contribute to the background pair
production. Interpolations between measurements of $\na$ and $\nb$ in
narrow centrality bins are used to estimate the average production of
single particles at any particular value of $N_{part}$. 

In a computational algorithm, $\xi$ can be calculated by throwing an $N_{part}$
value, according to the distribution of events within the centrality
selection, $w^{glaub}$, as taken from the Glauber Monte Carlo
calculation. The average number of 
type $A$ and $B$ particles is determined from the interpolated
centrality dependence and their product gives the number of
combinatorial pairs in the event. Events are created in this manner
until $\xi$ is numerically stable.  

The production of $A$ and $B$ particles at a given $N_{part}$ is
typically modelled with a Poisson distribution. However, the details
of this functional form do not affect the calculation so long as the
displacement from the average value is independent between triggers
and partners. To demonstrate this, $\xi$ has been calculated for a
delta function, a step function spanning $\pm25\%$ the average, and a pair
of asymmetric delta functions where the production at $-25\%$ was
twice that at $+50\%$ the average. The $\xi$ at 50-60\% centrality 
using these distributions was
found to be 1.1012(1), 1.1012(2), and 1.1013(2), respectively, where
the Poisson form gives 1.1010(6). This agreement is within the
statistical precision of the computational tests. Thus, even though
Poisson distributions are often used, they are not in
general required so long as the deviations from average are
independent.     

Using the insight that only the average value is relevant, $\xi$ can
be calculated equivalently from the Glauber distributions and the
yield interpolations by summing over all $N_{part}$ values. The
correction for centrality selection becomes a simple matter of finding
the event-weighted averages of the three functions 
($n^A$, $n^B$ and $n^A x n^B$) for the
centrality bin in question.  The expression for $\xi$ is re-written as:
\begin{eqnarray}
\label{eqn:xisums}
  \xi = \frac{\underset{i}{\sum} {n^A_i n^B_i w^{glaub}_{i}}}
  {\underset{i}{\sum} {n^A_i w^{glaub}_{i}}
    \underset{i}{\sum} {n^B_i w^{glaub}_{i}}}
  \underset{i}{\sum} {w^{glaub}_{i}}
\end{eqnarray}
where $i$ indexes sequentially over all $N_{part}$ values from 2 to
$N^{max}_{part}$ and $n^A_i = n^A(N^i_{part})$.

For trends of $n^A$ and $n^B$ that rise (or fall) in concert, the
value of $\xi$ will be always larger than 1.  If either $n^A$ or $n^B$
is independent of centrality, the correction is precisely 1.  If for
some reason, one trend rises and the other falls, $\xi$ will be less
than one.  In practice, the trends of trigger and partner production
rates with centrality are in the same direction and $\xi$ is an upward
correction on the production rate product. The magnitude of the
correction depends on how strongly the trends vary across the
centrality bin compared to the yield of the bin.  Since particle
production rises most quickly in peripheral events, the magnitude of
$\xi$ is largest in this region. For the same reason wider centrality
bins require larger corrections than more narrow bins. 

We now provide an example of calculating $\xi$ by using the charged
hadron yields published in Ref.~\cite{ppg023}. In practice,
uncorrected yields should be used to determine $\xi$ in order to
properly take into account the multiplicity dependence of the
reconstruction efficiency. Under all but extreme cases, the physical
centrality dependence dominates the value of $\xi$ as detector
efficiency usually varies only slowly with centrality. Therefore the
$\xi$ trends produced here contain the general features of a typical
calculation. 

\begin{figure}[ptb]
\centering
\begin{minipage}{1.0\linewidth}
\includegraphics[width=0.75\linewidth]{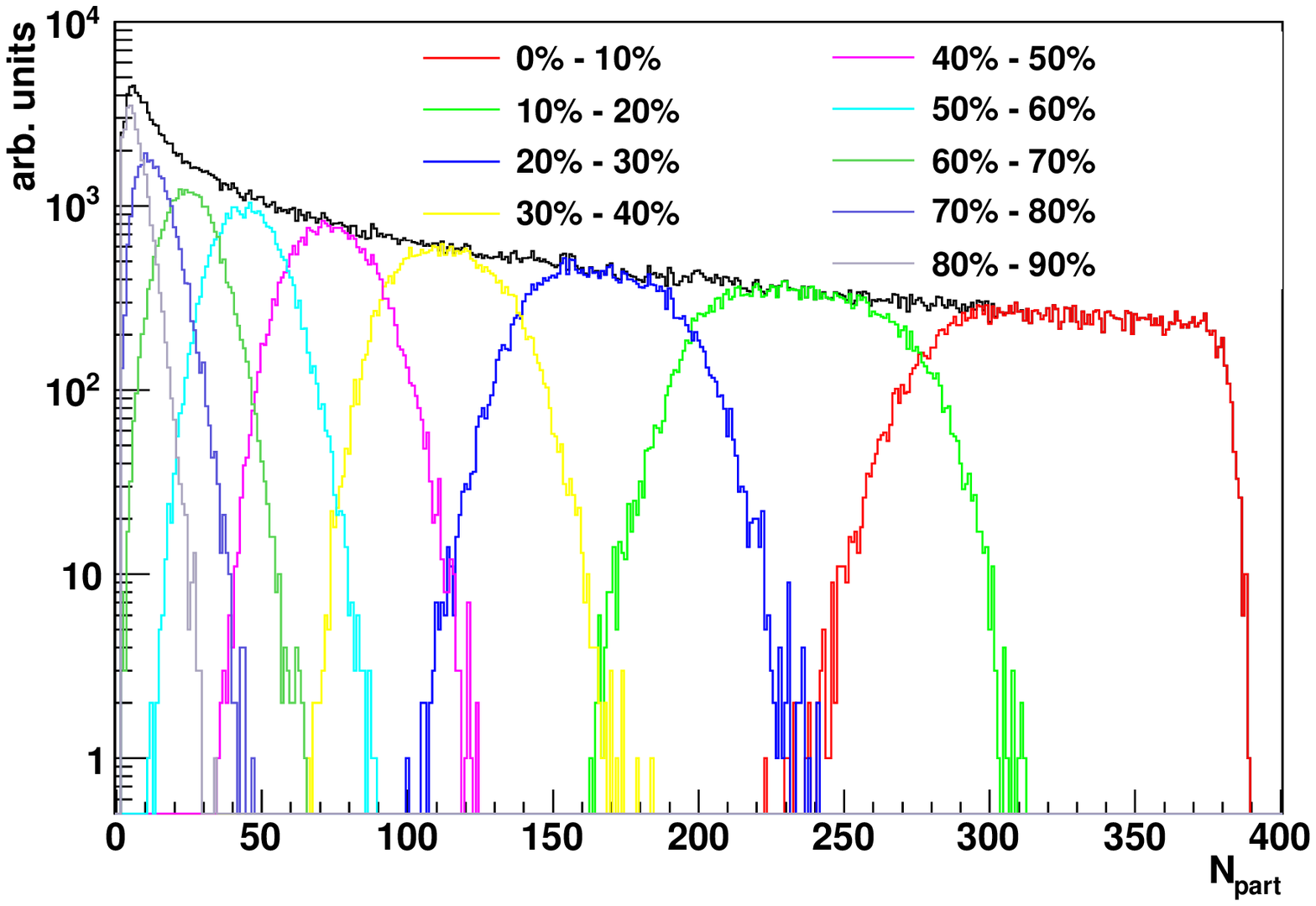}
\end{minipage}
\begin{minipage}{1.0\linewidth}
\includegraphics[width=0.75\linewidth]{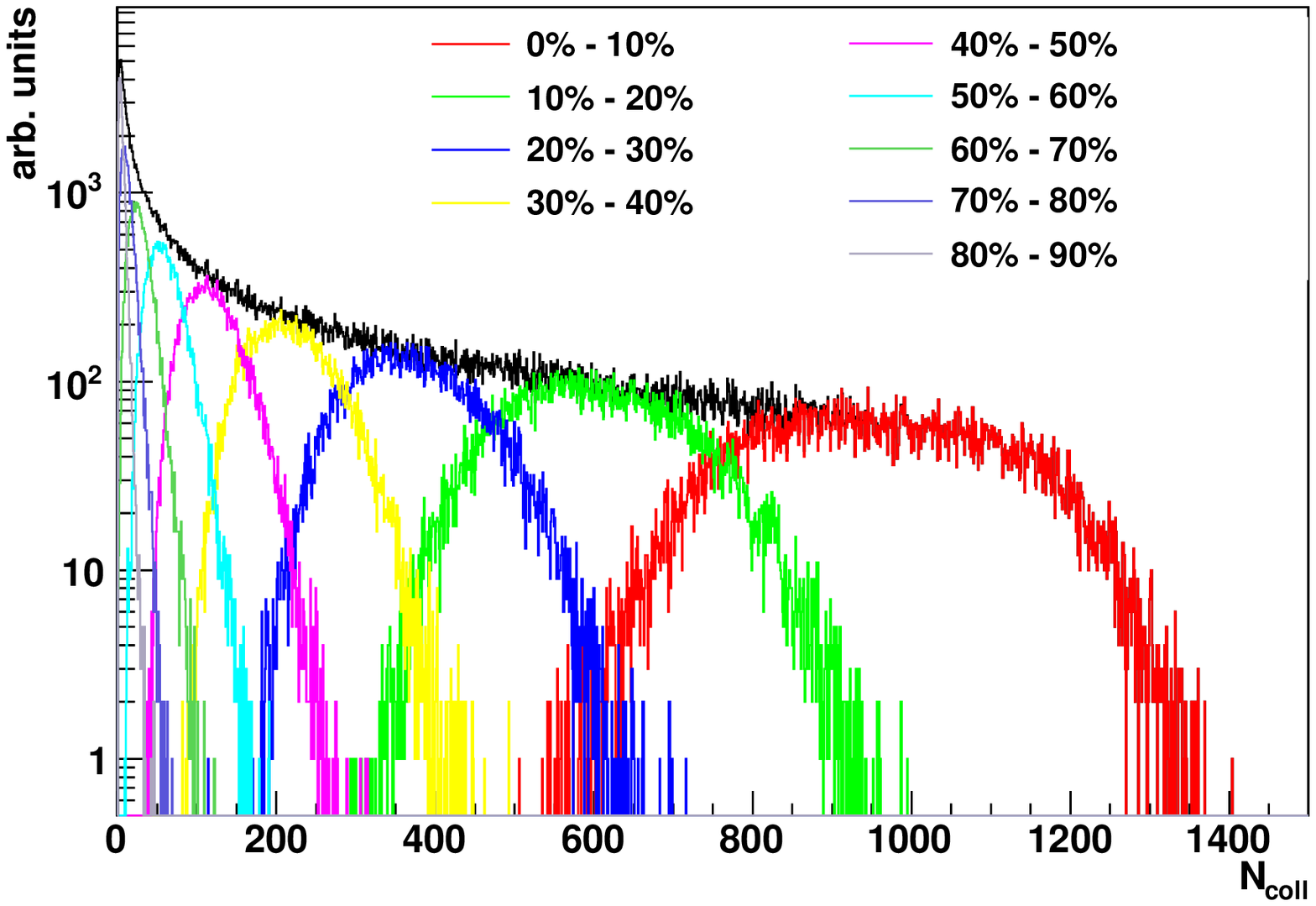}
\end{minipage}
\caption{(Color online) $N_{part}$ (top) and $N_{coll}$ (bottom) distributions from
  the Glauber Monte Carlo.}  
\label{figglabuberdistribution}
\end{figure}

\begin{figure}[phtb]
\centering
\begin{minipage}{1.0\linewidth}
\includegraphics[width=0.75\linewidth]{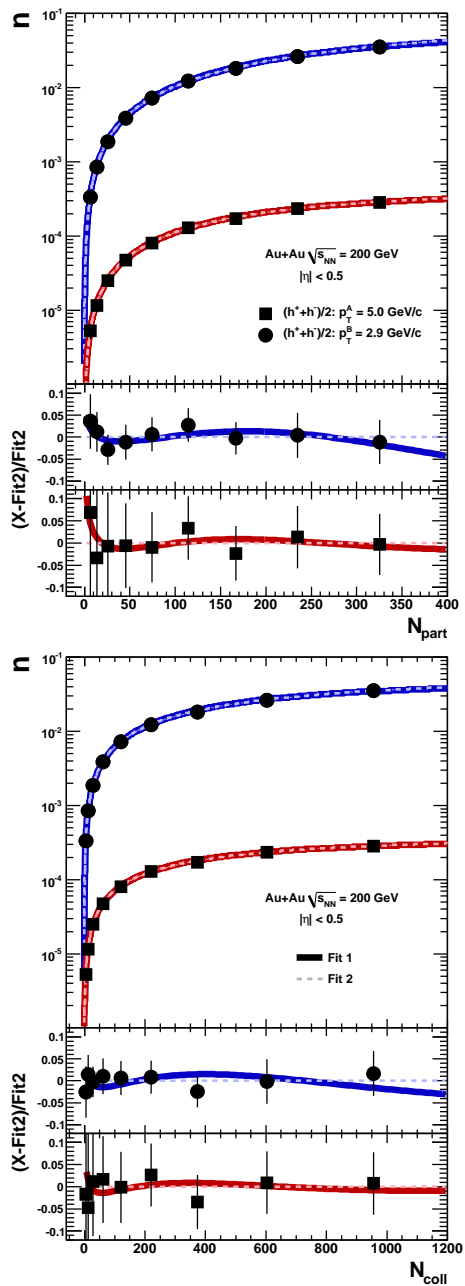}
\end{minipage}
\caption{(Color online) Invariant yield of charged hadrons as a function of
  $N_{part}$ and $N_{coll}$ at $p_T = 2.9$ GeV$/c$ and $p_T=
  5.0$ GeV$/c$. Data are from Ref.~\cite{ppg023} and the errors shown
  are statistical and centrality dependent systematic
  uncertainties. Fits are to Eq~\ref{eqtan} and Eq~\ref{eqsatexp}.}
\label{fits}
\end{figure}

Fig~\ref{figglabuberdistribution} shows the Glauber event distributions
for various centrality bin selections in both $N_{part}$ and
$N_{coll}$~\cite{glauber_review}. 
Invariant yields as a function of $N_{part}$ and $N_{coll}$ for
partners, $p^B_T=2.9$ GeV/$c$, and triggers, $p^A_T= 5.0$ GeV/$c$, are
shown in Fig~\ref{fits}. The data, $\na$ and $\nb$, are fit
with the following two functional forms, chosen for their smoothness
and well-controlled behavior for large $N$. The inverse tangent,
Eq~\ref{eqtan}, function is referred to as Fit 1 and the exponential
function, Eq~\ref{eqsatexp}, is denoted as Fit 2. 
\begin{eqnarray}
n^{\{A,B\}} = \gamma \arctan(\beta N^{\alpha})
\label{eqtan}
\end{eqnarray}
\begin{eqnarray}
n^{\{A,B\}} = \gamma ( 1 - e^{-\beta N^{\alpha}})
\label{eqsatexp}
\end{eqnarray}
where $N$ is either $N_{part}$ or $N_{coll}$. Sensitivity to the fit
functional form is assessed by comparison of the resulting $\xi$
values from use of the two fits.

The calculated values of $\xi$ from Eq~\ref{eqn:xisums} for these
trigger-partner selections are shown in Table~\ref{tabxi}. For central
collisions $\xi$ is a small correction to the background level,
however since the background level is large compared to the jet
signal, the effect of including the centrality correlations on the
extracted jet signal is substantial. The centrality correction
uncertainty is estimated from the spread in calculated values using
the $N_{part}$ vs. $N_{coll}$ description and from using the two choices of
functional forms.

\begin{table}[phtb]
\caption{$\xi$ values for charged hadrons pairs between 5.0 GeV$/c$ triggers and 2.9 GeV$/c$ partners.}
\begin{tabular}{|c|c|c|c|c|}
\hline
Centrality & $N_{part}$ & $N_{part}$ & $N_{coll}$ & $N_{coll}$ \\
(\%) & (Fit 1) & (Fit 2) & (Fit 1) & (Fit 2) \\
\hline
0-10  & 1.0041 & 1.0048 & 1.0041 & 1.0050 \\
10-20 & 1.0097 & 1.0107 & 1.0082 & 1.0089 \\
20-30 & 1.0205 & 1.0205 & 1.0150 & 1.0149 \\
30-40 & 1.0369 & 1.0353 & 1.0246 & 1.0236 \\
40-50 & 1.0606 & 1.0582 & 1.0405 & 1.0392 \\
50-60 & 1.1012 & 1.1005 & 1.0757 & 1.0753 \\
60-70 & 1.1825 & 1.1873 & 1.1604 & 1.1639 \\
70-80 & 1.2918 & 1.3065 & 1.2966 & 1.3091 \\
80-90 & 1.3952 & 1.4224 & 1.4419 & 1.4678 \\
\hline
\end{tabular}
\label{tabxi}
\end{table}

\subsection{Other Correlations}

The factorization of pair quantities into singles products appears
often in pair analyses and the degree to which the factorizations hold
should be examined in each case. 
Particle production rates were discussed above.  Quadrupole
anisotropies in general vary with centrality and are subject to 
an analogous centrality bias correction to Eq.~\ref{eq:v2fac}, though
in practice, it has been estimated to be a smaller effect than the 
uncertainties on the anisotropy values and neglected~\cite{ppg067}.
However, particle
production rates vary more significantly and the correction described
above is necessary. Additional correlations other than centrality
variation between triggers and partners could require additional
corrections of this kind. In practice, however, centrality binning correction is
the only significant correlation correction to background pair multiplicities. 

An example of another correlation of particle production rates
 is the position of the event along the
beam pipe within a detector with a finite acceptance on this
axis. Particle production in a symmetric collision system is peaked at
mid-rapidity and so more pairs will be reconstructed when the event is
centered with respect to the detector than when the event is
off-centered. Since the overall variation  
between the two types of events is small compared to the kinds of
variation seen above and can be made smaller with narrow binning, the
issue is much less of a problem than variations with centrality and
typically does not necessitate correction.

Another source could is triggers and
partners in the background are significantly produced via the decay of the
same parent particle. In this case, there is a correlation in the
number of background pairs related to the parent particle
multiplicity within the event. For the background in heavy ion
collisions, the source is mostly pairs from different production
centers and is unlikely to be strongly influenced by a single decay
process, however this should be checked with simulations based on the specific analysis
cuts and detectors used in a measurement.

\subsection{Pair Cuts}

Pair cuts are necessary in correlation analyses where
both trigger and partner particles are measured in the same detector
subsystem. If the trajectories of the trigger and partner lie
sufficiently close together within the detector, the deposited signals
may interfere constructively or destructively in their reconstruction
depending upon the nature of the reconstruction algorithms used. The
pair efficiency of these effects is extremely difficult to model to
the precision required in most analyses. Therefore pair efficiency is
simplified by cutting all pairs that fail proximity cuts in the
detector subsystems. With sufficient detector segmentation, only a small
region is masked and the percentage of pairs failing the cut will be
small. 

The influence on pair multiplicity due to rejection of pairs with
unacceptably small separation is quantified by $\kappa$, the fraction
of randomized pairs that survive the pair cuts. Some fraction of
random pairs will fail the cut due to the finite probability for two
particles to pass through the same region of the detector. This
probability can be estimated during event mixing. Since the masked
regions represent the same spatial coverage in all events, the value
of $\kappa$ has no observed dependence on event multiplicity.
The
multiplicity of the background, after taking into account losses due
to pair cuts, may be calculated from the singles distributions and
knowledge of the pair cut survival probability via 
\footnote{ 
  The value of $\kappa \na \na$ is equivalent to the mixed event pair
  multiplicity.}: 
\begin{eqnarray}
\label{kappa}
 \langle n^{AB}_{bg} \rangle = \xi \kappa \na \nb.
\end{eqnarray}

\subsection{Working Equation}

Thus, fully corrected ABS background levels in realistic scenarios may
be calculated in the manner described above as:
\begin{eqnarray}
  \label{b0_ABS_MSMP}
  b_{0} = \xi \frac{\kappa \na \nb}{\langle n^{AB}_{same} \rangle}.
\end{eqnarray}

\subsection{Limits of the Method}

The single particle production rate, $n^A$, can be written
as $n^A = j^A + b^A$ where $j^A$ are particles from jets and $b^A$ are
particles from non-jet sources. 
A similar decomposition can be made for type
$B$ particles. Using this notation, all pairs in the event can be 
expanded and factorized as: 
\begin{eqnarray}
  \langle n^A n^B \rangle 
  &=& \langle\left(\ja + \ba\right)\left(\jb + \bb\right)\rangle 
  \nonumber \\
  &=& \langle\ja\jb\rangle + \langle\ja\bb\rangle +
  \langle\jb\ba\rangle + \langle\ba\bb\rangle 
  \nonumber \\
  &=& \langle\ja\jb\rangle + \xi\kappa 
  \left[ \rule{0mm}{4.0mm} \right. 
  \langle\ja\rangle\langle\bb\rangle 
  \nonumber \\
  & & \hspace{7mm} + \langle\jb\rangle\langle\ba\rangle 
  + \langle\ba\rangle\langle\bb\rangle
  \left. \rule{0mm}{4.0mm} \right] 
\end{eqnarray}
The combinatorial background as estimated in ABS and expanded in terms
of $j$ and $b$ becomes: 
\begin{eqnarray}
\langle n^A \rangle \langle n^B \rangle &=& 
\langle j^A + b^A \rangle \langle j^B + b^B \rangle \nonumber \\
 & = & 
\langle j^A \rangle \langle j^B \rangle +
\langle j^A \rangle \langle b^B \rangle \nonumber \\
& + &  
\langle j^B \rangle \langle b^A \rangle +
\langle b^A \rangle \langle b^B \rangle 
\end{eqnarray}
Note that unlike the last three terms, the first term, $\langle j^A
\rangle \langle j^B \rangle$, is not part of the background. So the
ABS subtraction produces an extra term beyond the jet signal, $\langle
j^A j^B \rangle$ , such that:  
\begin{eqnarray}
\langle n^A n^B \rangle 
- \xi\kappa \langle n^A \rangle \langle n^B \rangle
= \langle j^A j^B \rangle 
- \xi \kappa \langle j^A \rangle \langle j^B \rangle
\end{eqnarray}
For the background subtraction to work without substantial
over subtraction, this extra term is required to be small with respect
to the jet signal.  
\begin{eqnarray}
\xi \kappa \langle j^A \rangle \langle j^B \rangle \ll \langle j^A j^B \rangle
\end{eqnarray}
Since hard scattering produces particles at rates determined by the 
characteristics of the scattering itself, like momentum transfer, the 
jet particle production rates for $A$ and $B$ particles will 
be highly correlated with one another. The presence of jet triggers 
increases the likelihood of production of jet partner particles within
the same event.

\section{The ZYAM method}

The Zero Yield At Minimum (ZYAM) methodology sets the normalization of
the background contribution through an assumption that the jet
contribution falls to zero yield at some point or region in
$\Delta\phi$. 

In addition to the assumption that the jet and background correlations
are from essentially independent sources and thus separable, the
validity of the ZYAM method is conditioned upon the existence of (a)
one or more points with vanishing yield in the actual jet
contribution, and (b) a sufficiently well-sampled correlation function
that enables a stable and precise determination of the minimum value.

In heavy ion collisions at sufficiently high transverse momenta
($p^{\{A,B\}}_T\gtrsim4$ GeV$/c$) or in $p$+$p$ collisions, the jet
contribution to the correlation function consists of well separated
near-side and away-side peaks~\cite{stardijet}. In these cases due to
the relative narrowness of the jet peaks, there is a broad region over
which the background contribution dominates and can be determined with
little bias under the ZYAM method. In the case of modified shapes in
the jet contribution such as those found at intermediate $p_{T}$ in
central heavy ion collisions, the ZYAM assumption could be broken by
jet broadening or other modifications creating yield between the
locations of the near- and away-side peaks. Without independent
verification of the ZYAM assumption, the method can potentially
produce unreliable results due to over-subtraction in these cases.

In the simplest, and least reliable, application of the ZYAM
procedure, the level of the background contribution is adjusted (with the
harmonic amplitude remaining fixed at its measured value) until one
measured bin in the jet function is zeroed. Clearly, small bins
relative to features in the jet contribution are needed to limit jet
contamination of the ZYAM bin. However, division of a fixed sample
size into smaller and smaller $\Delta\phi$ bins increases statistical
scattering, and hence the degree to which the lowest $\Delta\phi$ bin
is influenced by downward fluctuations. A slightly more sophisticated
method uses the average of three neighboring bins in place of a single
bin. The moving average of neighboring bins attempts to balance the
effects of jet contamination and statistical fluctuations (however,
depending on the width of the bins and the physics of interest, this
broader ZYAM region could make the assumption of a zero yield region much
less valid).  The most stable determination of background is to fit
the correlation function and raise the background contribution to touch the
fit at the minimum value. Assuming a reliable interpolation can be
found (which requires sufficient statistics or outside assumptions),
this method affords the best reliability against downward statistical
fluctuations.

The statistical uncertainty propagated from the ZYAM method can be
calculated with a simple Monte Carlo algorithm. The procedure
generates simulated correlation functions by sampling, bin by bin in
$\df$, a new point from a Gaussian whose mean and width are given by
the measured value and its uncertainty. The ZYAM procedure is then
performed on the generated distribution and a ZYAM background
normalization is extracted. The statistical uncertainty of the ZYAM
method is thus estimated by the variation of the background level over
multiple repetitions of this procedure. Such a calculation
demonstrates that simply taking the statistical uncertainty of the one
or three points in the first two implementations leads to an
underestimation of the statistical uncertainty, since it does not
account for any positional shift in $\df$ of the ZYAM point.

\begin{figure}[tb]
\begin{minipage}{1.0\linewidth} 
  \includegraphics[width=0.75\linewidth]{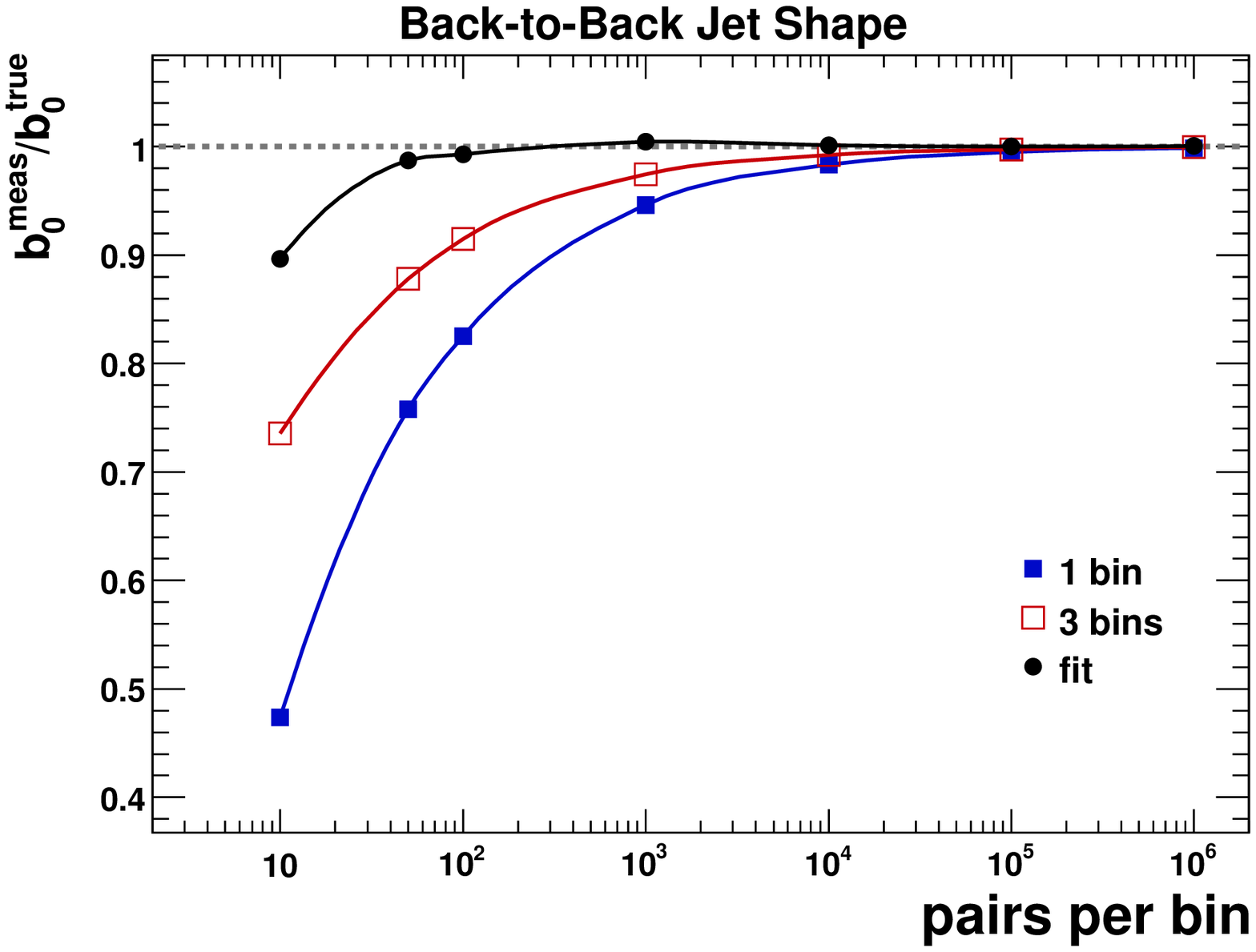}
\end{minipage}
\begin{minipage}{1.0\linewidth}
  \includegraphics[width=0.75\linewidth]{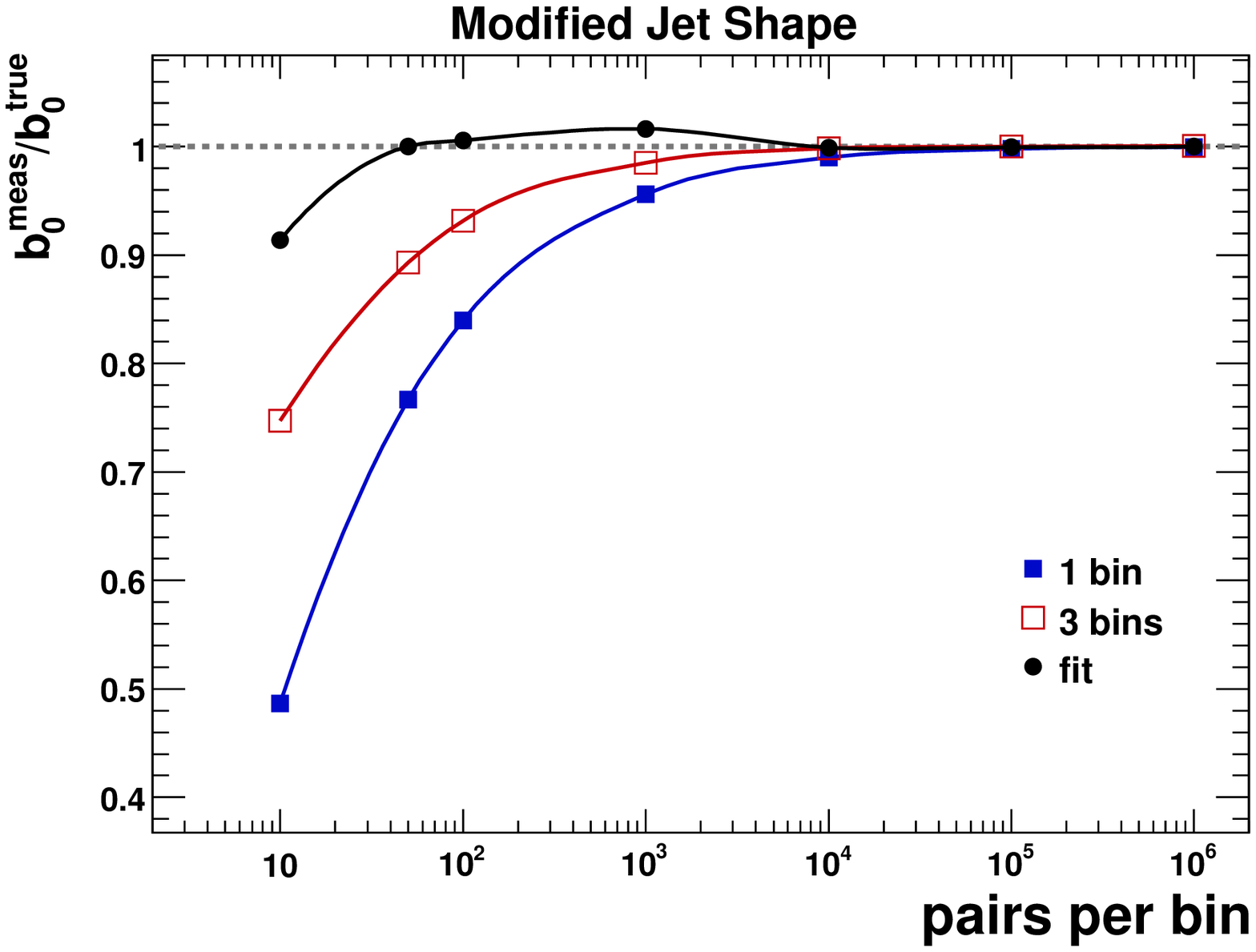}
\end{minipage}
\caption{(Color online) Crosschecks on ZYAM under subtraction due to statistical
  fluctuations for both back-to-back and modified jet shapes}
\label{zyamSys}
\end{figure}

Some failures in the ZYAM method can be tested by using a known
distribution under various levels of statistical sampling. There is a
strong downward tendency in the ZYAM procedure that must be carefully
avoided in order to extract reliable per-trigger jet pair yield
estimates. The procedure here is similar to that in extracting the
statistical scatter only it is the average offset from the true value
that is being examined. Clearly the true value in the measured
distribution is not known. However, a mock distribution similar to the
measured distribution can be asserted and then tested under
statistical samplings similar to the data. We show the results of two
such tests of functional forms. One distribution was chosen to follow
a back-to-back shape. The other was given an offset away-side peak. Both
results appear in Fig~\ref{zyamSys}.  
The choice of functional form does not significantly alter the resulting
under subtractions for the two cases tested. The jet to background
ratio was not varied in these tests. The single-bin implementation
of ZYAM is shown to be extremely sensitive to catching sizable statistical
fluctuations at low statistics. The three-bin method is more robust
against fluctuations, but also fails badly. The functional fit ZYAM method
works best, but is not completely robust against failure without the addition
of sensible constrains on the fits to the correlation functions.
No constraints, such as reasonable jet widths, were required
in these tests. Unreasonably narrow jet widths (less than one $\Delta\phi$
bin) are responsible for the failures in these fits at low statistics.

To summarize, the ZYAM method may fail in two cases. The method may
over subtract if there is a significant amount of jet yield at the 
ZYAM point. The method also has problems in some implementations with
under subtraction when applied at low statistics.  Given the
sensitivity of the ZYAM method to low statistics or extremely modified
jets, it is good practice to confirm the results independently with
the ABS method in these cases.  

\section{The Underlying Event}
\label{sec:ue}

The ABS background levels in the most peripheral bins have been found to lie
below the ZYAM background~\cite{ppg067}. Qualitatively, this is
expected because the ZYAM assumption puts an upper limit on the
background level. However, it is possible that measurements are also
sensitive to the underlying event, as seen in $p$+$p$
collisions~\cite{hep_ue}. The underlying event is thought to be
composed of initial and final state radiation as well as soft parton
interactions besides the one that created the observed jet. These
multi-parton interactions are not entirely uncorrelated with the
jet. Furthermore, as the background in a small system is the result of
very few soft parton interactions, the multiplicity resulting from a
single soft interaction to both trigger and partner may become an important
effect to model. Thus these effects may introduce additional
correlations in the background beyond the centrality correlations
which are removed by $\xi$.    

In large systems, where the background multiplicity is almost entirely
driven by impact parameter, these variations in the combinatorial
background play a much smaller role in the average background
multiplicities. Here the difference between ZYAM and the ABS
background can bracket the uncertainty on the background
subtraction. The ABS method will underestimate the background by not
including any underlying event and ZYAM will overestimate the
background, possibly removing some jet signal. In the small systems at
higher momenta, even this extreme in physics assumptions
translates into a small uncertainty on the extracted conditional
yields. However, small systems at lower momenta fair less well and
subtractions may produce significant uncertainties in the extracted
conditional yields. 

\section{Summary}

The separation of pairs originating in the underlying event from those
associated with jet production is of great importance for two particle
correlations in heavy ion collisions. We have provided the first
complete description of how to calculate the effect of centrality
correlations, which enables the combinatorial background to 
be subtracted without assumptions about the shape of the jet
correlations. This method has the additional advantage of having small
uncertainties, especially in the case of statistics limited probes
where the uncertainty on the ZYAM background will be large. This
method could trivially be generalized to three or more particle
correlations. Proper, consistent treatment of the background is
crucial to the quantitative extraction of properties of the hot dense
nuclear matter and its geometry from multi-particle correlation
measurements.

We have described a method for subtraction of the combinatorial
background under the assumption that there are two independent sources
of pairs. Reality, however is likely more complicated. The discussion
of the underlying event in Section~\ref{sec:ue} contains one example
of additional correlations which could be important. Currently, the
two source assumption has proven useful in the interpretation of two
particle correlation measurements. More precise measurements and more
complicated experimental questions could necessitate a re-evaluation
of the two source model. In the meantime, it is recommended that
publications include full correlation functions in order to make
available the data containing no assumptions about jet-background
decomposition should these considerations prove necessary.

\section{Acknowledgments}
We thank Tom Hemmick with whom some of this work was started, Paul
Stankus for the valuable discussions, and Dave Morrison for his helpful
comments. 

A. Adare acknowledges funding from the Division of Nuclear Physics of
the U.S. Department of Energy under Grant No. DE-FG02-00ER41152.
A. Sickles is supported by the U.S. Department of Energy under
contract DE-AC02-98CH1-886. 
M. McCumber is supported by the U.S. Department of Energy under Grant
No. DE-FG02-96ER40988.
\bibliography{abs}

\end{document}